\documentclass[aps,pre,showpacs,showkeys,amsmath,amsfonts,amssymb,superscriptaddress]{revtex4} 
\usepackage[dvips]{graphicx}
\usepackage{bbm}
\newcommand{\beq}{\begin{equation}} 
\newcommand{\eeq}{\end{equation}}
\newcommand{\bea}{\begin{eqnarray}} 
\newcommand{\eea}{\end{eqnarray}}

\input epsf 
 
\begin{document} 
 
\title{Density of states of continuous and discrete spin models: a case study} 

\author{Cesare Nardini} 
\email{cesare.nardini@gmail.com} 
\affiliation{Dipartimento di Fisica e Astronomia and Centro per lo Studio
delle Dinamiche Complesse (CSDC), Universit\`a di 
Firenze, via G.~Sansone 1, I-50019 Sesto Fiorentino (FI), Italy}  
\affiliation{Istituto Nazionale di Fisica Nucleare (INFN), Sezione di
Firenze, via G.~Sansone 1, I-50019 Sesto Fiorentino (FI), Italy}  
\affiliation{Laboratoire de Physique, \'Ecole Normale Sup\'erieure de Lyon, 46, all\'ee d'Italie, F-69007 Lyon, France}  
\author{Rachele Nerattini} 
\email{rachele.nerattini@unifi.it} 
\affiliation{Dipartimento di Fisica e Astronomia and Centro per lo Studio
delle Dinamiche Complesse (CSDC), Universit\`a di 
Firenze, via G.~Sansone 1, I-50019 Sesto Fiorentino (FI), Italy}  
\affiliation{Istituto Nazionale di Fisica Nucleare (INFN), Sezione di
Firenze, via G.~Sansone 1, I-50019 Sesto Fiorentino (FI), Italy}  
\author{Lapo Casetti} 
\email{lapo.casetti@unifi.it} 
\affiliation{Dipartimento di Fisica e Astronomia and Centro per lo Studio
delle Dinamiche Complesse (CSDC), Universit\`a di 
Firenze, via G.~Sansone 1, I-50019 Sesto Fiorentino (FI), Italy}  
\affiliation{Istituto Nazionale di Fisica Nucleare (INFN), Sezione di
Firenze, via G.~Sansone 1, I-50019 Sesto Fiorentino (FI), Italy}  

\date{\today} 
 
\begin{abstract} 
A relation between $O(n)$ lattice spin models and Ising models defined on the same lattice was recently put forward [L.\ Casetti, C.\ Nardini, and R.\ Nerattini, Phys.\ Rev.\ Lett.\ {\bf 106}, 057208 (2011)]. Such a relation, inspired by an energy landscape analysis, implies that the density of states of an $O(n)$ spin model on a lattice can be effectively approximated, at least close to the phase transition, in terms of the density of states of an Ising model defined on the same lattice and with the same interactions. In the present paper we show that such a relation exactly holds, albeit in a slightly modified form, in the special cases of the mean-field $XY$ model and of the one-dimensional $XY$ model. We also discuss the possible consequences of this result for the general case.
\end{abstract} 
 
 
 
\maketitle 

\section{Introduction}
\label{sec_intro}

Simple models play a prominent role in theoretical physics and especially in statistical mechanics. Moreover, finding links, relations or mappings---either exact or approximate---between different models often allows a deeper understanding of the models themselves and of the physics they describe.

In a recent paper \cite{prl2011} a relation between the microcanonical densities of states of continuous and discrete spin models was conjectured. More precisely, it was suggested that the density of states of an $O(n)$ classical spin model on a lattice can be approximated in terms of the density of states of the corresponding Ising model, i.e., an Ising model defined on the same lattice and with the same interactions. Such a relation is suggested by an ``energy landscape'' approach \cite{Wales:book} to the microcanonical thermodynamics of these models, the key observation being that all the configurations of an Ising model on a lattice are stationary points of an  $O(n)$ model Hamiltonian defined on the same lattice with the same interactions, for any $n$. The relation between the densities of states can be written as 
\beq
\omega^{(n)}(\varepsilon) \approx  \omega^{(1)}(\varepsilon) \,g^{(n)}(\varepsilon)\, ,
\label{omega_appr}
\eeq 
where $\varepsilon$ is the energy density, i.e., $\varepsilon = E/N$ where $E$ is the total energy and $N$ the number of spins, $\omega^{(n)}$ is the density of states of the $O(n)$ model, $\omega^{(1)}$ the density of states of the corresponding Ising model and $g^{(n)}$ is a function representing the volume of a neighborhood of the Ising configuration in the phase space of the $O(n)$ model. The function $g^{(n)}$ is unknown in general, but since it comes from local integrals over a neighborhood of the phase space one expects it is regular. The relation (\ref{omega_appr}) will be more precisely discussed in Sec.\ \ref{sec_dos}; it is an approximate one and the approximations involved are not easily controlled in general. As discussed in \cite{prl2011}, were this relation exact there would be a very interesting consequence: the critical energy densities of the phase transitions of all the $O(n)$ models on a given lattice would be the same and equal to that of the corresponding Ising model. Despite the fact that the relation is approximate\footnote{The relation cannot be exact at least in the form proposed in \protect\cite{prl2011} because it would imply wrong values (although with the correct sign) for the critical exponents; see also the discussion in Sec.\ \ref{sec_conclusions} of the present paper.}, according to available analytical and numerical calculations the critical energy densities are indeed, if not equal, very close to each other, whenever a phase transition is known to take place, at least for ferromagnetic models on $d$-dimensional hypercubic lattices. More precisely, the critical energy densities are the same and equal to the Ising one for all the $O(n)$ models with long-range interactions, as shown by the exact solution \cite{CampaGiansantiMoroni:jpa2003}, and the same happens for all the $O(n)$ models on a one-dimensional lattice with nearest-neighbor interactions. As far as $O(n)$ models with nearest-neighbor interactions on cubic lattices with $d > 1$ are concerned, only numerical results are available. According to available data, the transition energy densities are consistent between Ising, $O(2)$ and $O(3)$ models on $d=3$ cubic lattices, while the critical energies of the ferromagnetic transition of the Ising model and of the Bere\v{z}inskij-Kosterlitz-Thouless (BKT) transition of the $XY$ model on a two-dimensional lattice appear to be only slightly different, the difference being about 2\% (see Ref.\ \cite{prl2011} and references therein). At present, it is difficult to say whether transition energies are really equal for all cases but the BKT one or they are equal only for the long-range and the 1-$d$ cases and different in all the other cases, with so small a difference that it is not masked by statistical errors only in the BKT case. However, the above mentioned results show that assuming the validity of the relation as far as the location of the critical energy is concerned---i.e., assuming Eq.\ (\ref{omega_appr}) is a reasonable approximation close to the transition energy---gives in general a good prediction for the critical energy itself, and an exact prediction for the long-range and the 1-$d$ models. It must be stressed that the latter cases are very special, in that the critical energy density of the transition equals one of the boundaries of the energy density domain: the lower bound $\varepsilon_{\text{min}}$ in the 1-$d$ case and the upper bound $\varepsilon_{\text{max}}$ in the long-range case. 

Clearly, the fact that a particular prediction made using Eq.\ (\ref{omega_appr}) turns out to be exact does not imply that the equation itself is exact. This notwithstanding, it is reasonable to try to understand if in some of the cases where it gives the correct prediction for the critical energy the relation (\ref{omega_appr}) can be derived with a lesser degree of approximation, or even exactly. The aim of the present paper is to show that at least in two particular cases it can be done. The two cases that will be considered are the mean-field $XY$ model and the 1-$d$ nearest-neighbor $XY$ model, i.e., two $n=2$ representatives of the two classes of $O(n)$ models where the critical energies are known to be exactly equal to that of the corresponding Ising model. We will show that for these two models an expression very similar to Eq.\ (\ref{omega_appr}), and which reduces to Eq.\ (\ref{omega_appr}) when $\varepsilon \to \varepsilon_c$, can be derived exactly in the thermodynamic limit. The technical aspects of the derivation strongly rely on the peculiarities of the two models so that we do not see an immediate possibility of generalization of the results derived in this paper to generic $O(n)$ models. This notwithstanding, we are convinced that our derivation and results may help in understanding more deeply the relation between $O(n)$ and Ising models, as we shall argue at the end of the paper. 

The paper is organized as follows. In Sec.\ \ref{sec_dos} the stationary points approach and the approximations introduced in \cite{prl2011} leading to Eq.\ (\ref{omega_appr}) are recalled and discussed. Secs.\ \ref{sec_mfXY} and \ref{sec_1dXY} are devoted to the explicit derivation of the relation between the Ising model density of states and the density of states of the mean-field $XY$ and 1-$d$ $XY$ models, respectively. In Sec.\ \ref{sec_conclusions} the results are discussed in a more general perspective, with emphasis on generalization to general $d$ dimensional lattices.

\section{Stationary points and density of states}
\label{sec_dos}

Let us now recall the derivation of Eq.\ (\ref{omega_appr}) made in Ref.\ \cite{prl2011}. As already mentioned in Sec.\ \ref{sec_intro}, the approach is an ``energy landscape'' one, i.e., it is based on the study of the stationary points\footnote{A stationary point of a function $f$ is a point $p$ such as $df(p) = 0$.} of the Hamiltonian. The importance of stationary points in the study of the microcanonical thermodynamics of a system with Hamiltonian $\mathcal{H}$ can be understood in an intuitive way as follows. The entropy density $s$ is defined as 
\beq
s(\varepsilon) = \frac{1}{N}\log \omega(\varepsilon)\, 
\eeq
where we have set Boltzmann's constant to unity, and we shall retain this setting throughout the paper. For a system with $N$ degrees of freedom described by continuous variables the density of states $\omega$  can be written as 
\beq
\omega(\varepsilon) = \int_{\Gamma} \delta({\cal H} - N\varepsilon) \, d\Gamma = \int_{\Gamma \cap \Sigma_{\varepsilon}} \frac{d\Sigma}{\left|\nabla {\cal H}\right|}\, ,
\label{coarea}
\eeq
where $\Gamma$ is the phase space and $d\Gamma$ its volume measure, $\Sigma_{\varepsilon}$ is the hypersurface of constant energy $E = N\varepsilon$, and $d\Sigma$ stands for the $N-1$-dimensional Hausdorff measure. The rightmost integral stems from a coarea formula \cite{Federer:book}. At a stationary point, $\nabla{\cal H}=0$ and the integrand diverges, so that its contribution to $\omega$ is clearly important\footnote{It may be shown that the density of states is nonanalytic in correspondence with stationary points although such nonanlyticities become weaker as $N$ grows  \protect\cite{KSS:jstat2008,jstat2009} so that they cannot be generically associated with thermodynamic phase transitions. However a general relation between stationary points and phase transitions is believed to exist and has been discussed in \protect\cite{physrep2000,Kastner:rmp2008,KastnerSchnetz:prl2008,FranzosiPettini:prl,FranzosiPettini:npb,FranzosiPettiniSpinelli:npb} and references therein, although the problem is still an open one.}. In the following we shall assume that the stationary points contribution to $\omega$, or more precisely the contribution of a special class of stationary points, is indeed the most important one.

Let us consider a classical isotropic spin model defined on a lattice (or more generally on a graph) with Hamiltonian
\beq
{\cal H}^{(n)} = - \sum_{i,j=1}^N J_{ij} S_i \cdot S_j= - \sum_{i,j=1}^N J_{ij} \sum_{a = 1}^n S^a_i S^a_j~,
\label{H}
\eeq
where $i$ and $j$ run over the $N$ lattice sites and the classical spin vectors $S_i = (S_i^1,\ldots,S_i^n)$ have unitary norm, i.e., $\sum_{a = 1}^n \left( S^a_i \right) ^2 = 1$ $\forall i = 1,\ldots,N$. The real matrix $J_{ij}$ dictates the interactions; in case they are long-ranged a normalization is understood such as to obtain an extensive energy, using e.g.\ the Kac prescription \cite{CampaEtAl:physrep}. The Hamiltonian (\ref{H}) is globally invariant under the $O(n)$ group. In the special cases $n=1$, $n=2$, and $n=3$, one obtains the Ising, $XY$, and Heisenberg models, respectively.
The case $n=1$ is even more special because $O(1) \equiv \mathbb{Z}_2$ is a discrete symmetry group. In this special case the Hamiltonian (\ref{H}) becomes the Ising Hamiltonian 
\beq
{\cal H}^{(1)} = - \sum_{i,j=1}^N J_{ij} \sigma_i \sigma_j~,
\label{H_1}
\eeq
where $\sigma_i = \pm 1$ $\forall i$. In all the other cases $n \geq 2$ the $O(n)$ group is continuous; each spin vector $S_i$ lives on an $n-1$ unit sphere $\mathbb{S}_1^{n-1}$. 

Let us now consider the stationary configurations of ${\cal H}^{(n)}$ for $n\geq 2$, i.e., the solutions $\overline{S} = (\overline{S}_1,\ldots, \overline{S}_N)$ of the $N$ vector equations 
\beq
\nabla {\cal H}^{(n)} = 0\,,
\label{eq_statpoints}
\eeq
with the constraint $\sum_{a=1}^{N}(S_{i}^{a})^{2}$.

It is impossible to find explicitly all the stationary points (except for some special case, see e.g.\ \cite{MehtaKastner:annphys2011}, of for very small systems, see e.g.\ \cite{Mehta:pre2011} and references therein). However, as shown in \cite{prl2011}, a particular class of solutions can be found by assuming that all the spins are parallel or antiparallel: $S^1_i = \cdots = S^{n-1}_i = 0$ $\forall i$. In this case the constraints $(S^n_i)^2 = 1$ imply $S^n_i = \sigma_i$ $\forall i$, and one finds that the stationary points equations (\ref{eq_statpoints}) are satisfied by any of the $2^N$ possible choices of the $\sigma$'s. The Hamiltonian (\ref{H}) becomes the Ising Hamiltonian (\ref{H_1}) when the spins belong to this class of stationary configurations. Therefore we have a one-to-one correspondence between a class of stationary configurations of the Hamiltonian (\ref{H}) of a $O(n)$ spin model and all the configurations of the Ising model (\ref{H_1}), i.e., the Ising model defined on the same graph with the same interaction matrix $J_{ij}$; the corresponding stationary values are just the energy levels of this Ising Hamiltonian. We shall refer to the class of stationary configurations $\overline{S}_i = (0,\ldots,0,\sigma_i) ~  \forall i = 1,\ldots,N$ as ``Ising stationary configurations''. 
There will be also other stationary configurations; nonetheless, the $2^N$ Ising ones are a non-negligible fraction of the whole, especially at large $N$ because the number of stationary points of a generic function of $N$ variables is expected to be ${\cal O}(e^N)$ \cite{MehtaKastner:annphys2011,Schilling:physicad2006}. 

The above results hold for $O(n)$ and Ising models defined on any graph. From now on we shall restrict to regular $d$-dimensional hypercubic lattices and to ferromagnetic interactions $J_{ij} > 0$. In this case, in the thermodynamic limit $N\to \infty$ the energy density levels of the Ising Hamiltonian (\ref{H_1}), ${\cal H}^{(1)}(\sigma_1,\ldots,\sigma_N)/N ~ \forall \sigma_i = \pm 1$, become dense and cover the whole energy density range of all the $O(n)$ models. This fact, together with the above mentioned fact that their number is exponentially large in $N$, suggests that Ising stationary configurations are the most important ones, so that we may approximate the density of states $\omega^{(n)}(\varepsilon)$ of an $O(n)$ model in terms of these configurations. To this end, let us first rewrite the density of states of our $O(n)$ spin model as a sum of integrals over a partition of the phase space:  
\beq
\omega^{(n)}(\varepsilon) = \sum_p \int_{U_p} \delta({\cal H}^{(n)} - N\varepsilon) \, d\Gamma = \sum_p \int_{U_p\cap\Sigma_{\varepsilon}} \frac{d\Sigma}{\left|\nabla {\cal H}^{(n)}\right|}\, ,
\label{coarea_part}
\eeq
where $p$ runs over the $2^N$ Ising stationary configurations and $U_p$ is a neighborhood of the $p$-th Ising configuration such that $\left\{U_p\right\}_{p=1}^{2^N}$ is a proper partition of the configuration space $\Gamma = \left(\mathbb{S}^{n-1}\right)^N$, that coincides with phase space for spin models (\ref{H}). Since Ising configurations are isolated points in the configuration space of a $O(n)$ model, such a partition always exists. 

Until now we did not introduce any approximation. In Ref.\ \cite{prl2011}, two approximations were introduced to derive Eq.\ (\ref{omega_appr}) from Eq.\ (\ref{coarea_part}). Let us review and discuss them, in view of their r\^{o}le in the explicit calculations we shall carry out in the next Sections.  
\begin{description}
\item[{$(i)$}] It was assumed that the integrals in Eq.\ (\ref{coarea_part}) depend only on $\varepsilon$, i.e., the neighborhoods $U$ can be chosen, or deformed, such as
\beq
\int_{U_{p}} \delta({\cal H}^{(n)} - N\varepsilon) \, d\Gamma = \int_{U_{q}} \delta({\cal H}^{(n)} - N\varepsilon) \, d\Gamma = g^{(n)}(\varepsilon)  
\label{g}
\eeq
for any $p,q$ such that ${\cal H}^{(n)}(p) = {\cal H}^{(n)}(q) = N\varepsilon$. 
\item[{$(ii)$}] At a given value of $\varepsilon$, the largest contribution to $\omega^{(n)}(\varepsilon)$ in Eq.\ (\ref{coarea_part}) is likely to come from those $U_p$ such that ${\cal H}^{(n)}(p) = N\varepsilon$, because if ${\cal H}^{(n)}(q) \not= N\varepsilon$ then $\left|\nabla {\cal H}^{(n)}(x)\right| \not= 0$ $\forall x \in U_q \cap \Sigma_\varepsilon$, unless a zero in $\left|\nabla {\cal H}^{(n)}(x)\right|$ comes from a stationary configuration which does not belong to the Ising class. Since it was assumed that non-Ising stationary configurations could be neglected, only neighborhoods centered around stationary configurations at energy density $\varepsilon$ have been retained in the sum (\ref{coarea_part}). 
\end{description}
As shown in Ref.\ \cite{prl2011}, these two assumptions\footnote{We note that in Ref.\ \protect\cite{prl2011} assumption $(i)$ was referred to as assumption $(ii)$ and viceversa. Here we choose this ordering for the sake of clarity.} immediately lead to Eq.\ (\ref{omega_appr}). 
Both assumptions are needed to arrive to Eq.\ (\ref{omega_appr}), and are strictly related to each other. However, these two assumptions might well play a very different r\^ole. As we shall see in the following sections, in the two analytically tractable special cases, assumption $(ii)$ does not hold in general: it holds only when $\varepsilon \to \varepsilon_c$. As a consequence, one has to include also stationary configurations with energy $\varepsilon' \not = \varepsilon$ in the sum. Clearly, if assumption $(ii)$ does not hold, also assumption $(i)$ is of little use as such, since also neighborhoods centered around stationary points with energy density different from $\varepsilon$ have to be included in the sum. 

One might then replace assumption $(i)$ with 
\begin{description}
\item[{$(i')$}] The integrals in Eq.\ (\ref{coarea_part}) depend only on $\varepsilon$ and on the energy density $\varepsilon'$ of the stationary point, i.e., 
\beq
\int_{U_{p}} \delta({\cal H}^{(n)} - N\varepsilon) \, d\Gamma =  G^{(n)}(\varepsilon,\varepsilon')\, ,  
\label{big_g}
\eeq
for any $p$ such that ${\cal H}^{(n)}(p) = N\varepsilon'$. 
\end{description}
The function $g^{(n)}(\varepsilon)$ would then be related to $G^{(n)}(\varepsilon,\varepsilon')$ by
\beq
g^{(n)}(\varepsilon) = G^{(n)}(\varepsilon,\varepsilon)\, .
\label{bigsmallg}
\eeq
Using assumption $(i')$ alone, without invoking\footnote{One may wonder whether the removal of assumption $(ii)$ has any consequence on the robustness of the hypothesis of dominance of the Ising configurations. In our opinion it does not have any consequence, because the latter hypothesis is preliminary to the others, and relies on that Ising configurations are exponentially large in $N$, as the total number of stationary points is expected to be, so that they are at least a non-negligible fraction of the whole.} assumption $(ii)$, one obtains from Eq.\ (\ref{coarea_part}) the following expression for the density of states of a $O(n)$ model:
\beq
\omega^{(n)}(\varepsilon) = \sum_{\varepsilon'} \omega^{(1)}(\varepsilon') \, G^{(n)}(\varepsilon,\varepsilon')\, ,
\label{omega_conv}
\eeq
i.e., a convolution between the Ising density of states $\omega^{(1)}$ and the function $G^{(n)}$. Then, in the thermodynamic limit $N\to\infty$ a saddle-point-like mechanism might single out a value $\tilde{\varepsilon}$ for $\varepsilon'$, so that the convolution (\ref{omega_conv}) becomes a product:
\beq
\omega^{(n)}(\varepsilon) = \omega^{(1)}(\tilde{\varepsilon}) \, G^{(n)}(\varepsilon,\tilde{\varepsilon})\, ,
\label{omega_prod}
\eeq
where $\tilde{\varepsilon}$ is a suitable function of $\varepsilon$. If $\tilde{\varepsilon} = \varepsilon$, then using Eq.\ (\ref{bigsmallg}) one recovers Eq.\ (\ref{omega_appr}).
This is precisely what happens when $\varepsilon \to \varepsilon_c$ in the two special cases we are going to discuss in the following sections. In sec.\ \ref{sec_conclusions}, we shall argue about the possible generality of this scenario.

\section{The mean-field $XY$ model}
\label{sec_mfXY}

We shall now show that the density of states of the mean-field $XY$ model can be written in the form (\ref{omega_prod}), with $\tilde{\varepsilon} \to \varepsilon$ when $\varepsilon \to \varepsilon_c$.

The mean-field $XY$ model \cite{AntoniRuffo:pre1995} is a system of $N$ globally coupled planar spins (or alternatively of $N$ globally interacting particles constrained on a ring), with Hamiltonian
\beq
\mathcal{H}_{\text{MF}} = - \frac{1}{2N}\sum_{i,j=1}^N \cos \left(\vartheta_i - \vartheta_j \right)\, ,
\eeq
where $\vartheta_i \in [0,2\pi)$, so that the configuration (or phase) space of the system is the torus $\mathbb{T}^N$. This model has a mean-field phase transition from a ferromagnetic (or clustered, if one thinks of particles) to a paramagnetic (or uniform) phase at $\varepsilon_c = \varepsilon_{\text{max}} = 0$ and has been thoroughly studied being one of the simplest models of systems with long-range interactions \cite{CampaEtAl:physrep}; it belongs to the class (\ref{H}) with $n=2$ and $J_{ij} = 1/N$.
By introducing the magnetization\footnote{With a certain abuse of language we denote by magnetization the phase space function whose statistical average is the magnetization.} density vector ${\bf m} = \left(m_x,m_y\right)$, where
\bea
m_x & = & \frac{1}{N} \sum_{i=1}^N \cos \vartheta_i\, ,\\ 
m_y & = & \frac{1}{N} \sum_{i=1}^N \sin \vartheta_i\, ,\\ 
\eea
we can write the total energy of the system as a function of the modulus $m = |{\bf m}|$ of the magnetization density:
\beq
\mathcal{H}_{\text{MF}} = - \frac{N m^2}{2} \, .
\label{hm}
\eeq
For $XY$ models, Ising stationary points are configurations where the angles $\vartheta_i$ differ from each other by either $0$ or $\pi$. Due to the $O(2)$ invariance of the Hamiltonian, these stationary solutions are not isolated but belong to a manifold. We make them isolated by fixing\footnote{This does not affect the thermodynamics of the system in the $N\to\infty$ limit but for the fact that it chooses the direction of the breaking of the $O(2)$ symmetry below the critical energy density in such a way that $\left\langle m_y\right\rangle \equiv 0$ also in the broken symmetry phase.} $\vartheta_N = 0$, so that the Ising stationary configurations are all the configurations $\overline{\vartheta} = \left\{\overline{\vartheta}_i\right\}_{i=1}^N$ where the angles are either $0$ or $\pi$, and can be parametrized by the number $N_\pi$ of angles equal to $\pi$. The configurations with given $N_\pi$ are
\bea
\overline{\vartheta}_i & = & \pi ~~~~ \forall\, i = 1,\ldots,N_\pi \\
\overline{\vartheta}_i & = & 0 ~~~~ \forall\, i = N_\pi + 1,\ldots,N 
\eea
and all the others obtained by permutations of the indices $i$. The number $\nu(N_\pi)$ of such configurations is given by the binomial coefficient
\beq
\nu(N_\pi) = \frac{N!}{N_{\pi}!(N-N_{\pi})!}\, ,
\label{nu}
\eeq
while their magnetization and energy density depend only on $N_\pi$ and are given by
\bea
m(N_\pi) & = & m_x(N_\pi)= \frac{N - 2N_\pi}{N} = 1 - 2n_\pi\, ,\\
\varepsilon(N_\pi) & = & - \frac{m^2(N_\pi)}{2} = - \frac{\left(N - 2N_\pi\right)^2}{2 N^2} = - \frac{(1 - 2n_\pi)^2}{2}\, , \label{enpi}
\eea
where we have introduced the fraction of angles equal to $\pi$, $n_\pi = N_\pi/N$.

Given a stationary configuration $p = \left\{\overline{\vartheta}_1,\ldots,\overline{\vartheta}_N \right\}$, let us define the neighborhood  
\beq
U_p = \left\{
\begin{array}{ccc}
{\displaystyle \vartheta_i \in \left[\frac{\pi}{2},\frac{3\pi}{2} \right]}  & \text{if} &   \overline{\vartheta}_i = \pi  \\
& & \\
{\displaystyle \vartheta_i \in \left[\frac{3\pi}{2},\frac{\pi}{2} \right]}  &\text{if}&    \overline{\vartheta}_i = 0  
\end{array} \right .
\label{partition}
\eeq
so that $\{ U_p \}_{p=1}^{2^N}$ is a partition of the phase space $\mathbb{T}^N$. The density of states $\omega_{\text{MF}}$ of the mean-field $XY$ model can thus be written as
\beq
\omega_{\text{MF}}(\varepsilon) = \sum_{N_\pi=0}^N \nu(N_\pi)\, G_{\text{MF}}(\varepsilon,N_\pi)\,
\label{omega_mf_conv}
\eeq
where
\beq
G_{\text{MF}}(\varepsilon,N_\pi) = \int_{\pi/2}^{3\pi/2}d\vartheta_1 \cdots d\vartheta_{N_\pi} \int_{3\pi/2}^{\pi/2}d\vartheta_{N_\pi + 1} \cdots d\vartheta_{N}\, \delta\left[\mathcal{H}_{\text{MF}}(\vartheta_1,\ldots,\vartheta_N) - N\varepsilon \right]~. 
\label{Ge}
\eeq
We note that $\nu(N_\pi)$ given by Eq.\ (\ref{nu}) is nothing but the density of states $\omega^{(1)}_{\text{MF}}$ of the mean-field Ising model
\beq
\mathcal{H}^{(1)}_{\text{MF}} = - \frac{1}{2N}\sum_{i,j=1}^N \sigma_i \sigma_j \, ,
\label{H_ising_mf}
\eeq
as a function of the number of ``up'' spins $\sigma = 1$; using the relation (\ref{enpi}) to obtain the energy density $\varepsilon'$ of the Ising stationary configuration as a function of $N_\pi$, Eq.\ (\ref{omega_mf_conv}) can be written as
\beq
\omega_{\text{MF}}(\varepsilon) = \sum_{\varepsilon'} \omega_{\text{MF}}^{(1)}(\varepsilon') \, G_{\text{MF}}(\varepsilon,\varepsilon')\, ,
\label{omega_mf_conv_eps}
\eeq
where the sum runs over the energy density levels of the Ising mean-field Hamiltonian (\ref{H_ising_mf}), so that it is exactly Eq.\ (\ref{omega_conv}) written in the special case of the mean-field $XY$ model. It is important to stress that this result is a consequence of the fact that the energy of a Ising stationary configuration depends only on $N_\pi$ and that all the neighborhoods $U_p$ with the same $N_\pi$ contribute equally to the sum (\ref{omega_mf_conv}).

Let us now compute the function $G_{\text{MF}}$ defined in Eq.\ (\ref{Ge}). To make the calculation simpler it is useful to express $G_{\text{MF}}$ as a function of $m$ instead of $\varepsilon$; one then gets back to $\varepsilon$ using Eq.\ (\ref{hm}). Since we fixed the magnetization to be along the $x$ axis, the function $G_{\text{MF}}(m,N_\pi)$ is given by
\begin{equation}\label{GmNpi}
G_{\text{MF}}(m,N_\pi)=\int_{\pi/2}^{3\pi/2}\,d\vartheta_1...d\vartheta_{N_{\pi}}\,\int_{3\pi/2}^{\pi/2}\,d\vartheta_{N_{\pi}+1}...d\vartheta_N \, \delta\left(\sum_{i=1}^N \cos \vartheta_i-Nm\right)\,\delta\left(\sum_{i=1}^N \sin\vartheta_i\right)\,.
\end{equation}
Using the integral representation of the Dirac delta distribution, Eq.\ (\ref{GmNpi}) becomes
\begin{align}\label{Gintegraldelta}
G_{\text{MF}}(m,N_\pi)=&\left(\frac{1}{2\pi}\right)^2\int_{\pi/2}^{3\pi/2}\,d\vartheta_1...d\vartheta_{N_{\pi}}\,\int_{3\pi/2}^{\pi/2}\,d\vartheta_{N_{\pi}+1}...d\vartheta_N \,\int_{-\infty}^{\infty}dq_1\,\int_{-\infty}^{\infty}dq_2 \, \nonumber \\&\exp\left[iq_1\left(\sum_{i=1}^N \cos \vartheta_i-Nm\right)\right]\,\exp\left[iq_2\left(\sum_{i=1}^N \sin\vartheta_i\right)\right]\,;
\end{align}
by writing 
\bea
A(q_1,q_2)&=&\int_{\pi/2}^{3\pi/2}d\vartheta\,\exp\left[iq_1\,\cos\vartheta+iq_2\,\sin\vartheta\right]\, , \label{A}\\
B(q_1,q_2)&=&\int_{3\pi/2}^{\pi/2}d\vartheta\,\exp\left[iq_1\,\cos\vartheta+iq_2\,\sin\vartheta\right]=\int_{\pi/2}^{3\pi/2}d\vartheta\,\exp\left[iq_1\,\cos(\vartheta-\pi)+iq_2\,\sin(\vartheta-\pi)\right] \, , \label{B}
\eea
we get
\beq
G_{\text{MF}}(m,N_\pi)=\left(\frac{1}{2\pi}\right)^2 \,\int_{-\infty}^{\infty}dq_1\,\int_{-\infty}^{\infty}dq_2 \,\exp\left[N\left(-i m q_1 + n_{\pi}\log A(q_1,q_2)+(1-n_{\pi})\log B(q_1,q_2)\right)\right]\,.
\label{Gsaddle}
\eeq

The integrals in Eq.\ (\ref{Gsaddle}) can be computed with the saddle-point method \cite{BenderOrszag:book} in the limit $N\to\infty$. The saddle point is given by  $q_2=0$ e $q_1=-i \gamma$, where $\gamma\in\mathbb{R}$ satifies the self-consistency equation
\begin{equation}\label{autoconsistenza_punto_sella}
m=n_{\pi}\,\frac{I_1(\gamma)-L_{-1}(\gamma)}{I_0(\gamma)-L_0(\gamma)} + (1-n_{\pi})\,\frac{I_1(\gamma)+L_{-1}(\gamma)}{I_0(\gamma)+L_0(\gamma)}\,;
\end{equation}
in Eq.\ (\ref{autoconsistenza_punto_sella}), $I_k(\gamma)$ are modified Bessel functions of order $k$ and $L_k(\gamma)$ are modified Struve functions of order $k$ \cite{AbramowitzStegun:book}. We can thus write, in the thermodynamic limit $N\to\infty$, 
\beq
G_{\text{MF}}(m,n_{\pi})=\left(\frac{1}{2\pi}\right)^2 \,\exp\left\{N\left[-m \gamma + n_{\pi}\log \tilde A(\gamma,0)+(1-n_{\pi})\log \tilde B(\gamma,0)\right]\right\}\,,
\label{GMFsaddle}
\eeq
where we have written $n_\pi$ instead of $N_\pi$ since we are in the $N\to\infty$ limit, $\gamma$ must be numerically determined solving Eq.\  (\ref{autoconsistenza_punto_sella}), and the functions $\tilde A$ and $\tilde B$ are
\bea
\tilde A(\gamma,0)&=&\pi[I_0(\gamma)-L_0(\gamma)]\, , \label{Agamma}\\
\tilde B(\gamma,0)&=&\pi[I_0(\gamma)+L_0(\gamma)]\,. \label{Bgamma}
\eea
We can thus write, in the large $N$ limit, the density of states (\ref{omega_mf_conv}) as a function of $m$ as
\beq
\omega_{\text{MF}}(m) = \int_{0}^1 dn_{\pi}\,\exp\left[N(-m\gamma+n_{\pi}\log \tilde A(\gamma,0)+(1-n_{\pi})\log \tilde B(\gamma,0)-n_{\pi}\log n_{\pi}-(1-n_{\pi})\log (1-n_{\pi}))\right]\,,
\label{dens_stati_xymf_finale_implicita}
\eeq
where we have neglected the subleading contributions in $N$. Again, the integral (\ref{dens_stati_xymf_finale_implicita}) can be computed with the saddle-point method as $N\to\infty$, so that, given $m$ and thus $\varepsilon$, only a particular value of $n_\pi$ (and thus of $m'$ and, in turn, of $\varepsilon'$) is singled out and the density of states $\omega_{\text{MF}}$ assumes the product form (\ref{omega_prod}). The particular value of $n_\pi$ which is singled out is the one such that the exponent in Eq. (\ref{dens_stati_xymf_finale_implicita}) is maximum; it has to be computed numerically. 

\begin{figure}
\centering
\includegraphics[scale=1]{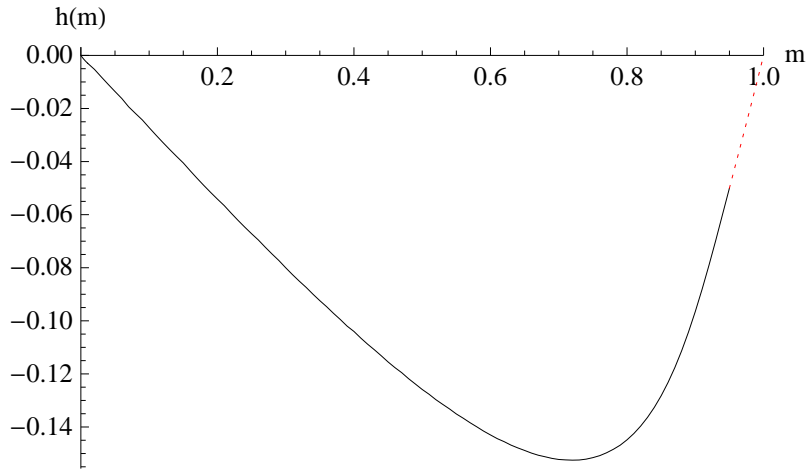} 
\caption{Numerical results for the function $h(m)$ defined in Eq.\ (\protect\ref{h}) for the mean-field $XY$ model. The red (dotted) part of the curve is obtained by interpolation (see text).}
\label{figure_h}
\end{figure}

The saddle point on Eq.\ (\ref{dens_stati_xymf_finale_implicita}) singles out a value $\tilde{m}$ of the magnetization such that 
\beq
\omega_{\text{MF}}(m) = \omega^{(1)}(\tilde{m}) \, G_{\text{MF}}(m,\tilde{m})\, . \label{omegaprod_m}
\eeq
In order to show that the value of $\tilde{m}$ as a function of $m$ converges to $m$ as $m \to m_c$, where $m_c = 0$ is the critical value of the magnetization, in Fig.\ \ref{figure_h} we plot the function 
\beq
h(m) = m - \tilde{m}\, .
\label{h} 
\eeq
Figure \ref{figure_h} shows that $h \to 0$ as $m \to 0$, so that the density of states $\omega_{\text{MF}}(m)$ is such that
\beq
\omega_{\text{MF}}(m) \to \omega^{(1)}({m}) \, g_{\text{MF}}(m)\, , \label{omegaprod_mc}
\eeq
where $g_{\text{MF}}(m) = G_{\text{MF}}(m,m)$, for $m \to m_c$. More precisely, $h$ appears to be a linear function of $m$ as $m \to 0$, $h(m)\propto - m$. 
When $m \to 1$ the numerical procedure we used to compute $h(m)$ had some convergence problems. Since $m=1$ implies $h(m)=0$ and $n_{\pi} = 1$, to avoid these numerical problems the curve plotted in Fig.\ \ref{figure_h} in the range $m\in\left[0.97,1\right]$ has been evaluated interpolating the numerical results obtained for $m<0.97$ with the constraint $h(1)=0$. The interpolating curve is drawn in red and in dotted style in Figure \ref{figure_h}. We stress that the part of the curve relevant to the phase transition is that in the opposite limit, $m \to 0$, where the numerical procedure easily converges.

\begin{figure}
\centering
\includegraphics[scale=1]{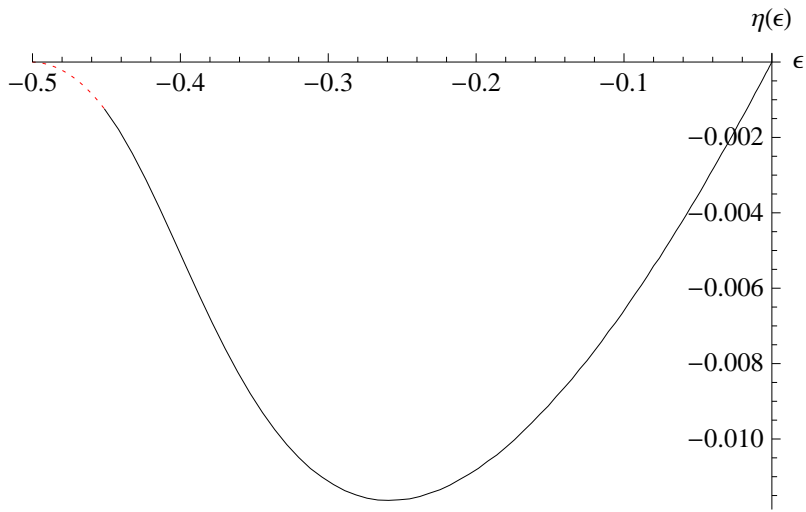} 
\caption{Numerical results for the function $\eta(\varepsilon) = h(-\sqrt{2\varepsilon})$ defined in Eq.\ (\protect\ref{eta}) for the mean-field $XY$ model. The red (dotted) part is obtained by interpolation (see Fig.\ \protect\ref{figure_h} and text).}
\label{figure_eta}
\end{figure}

We can now go back to the energy, using $\varepsilon = -m^2/2$, and write
\beq
\omega_{\text{MF}}(\varepsilon) = \omega^{(1)}(\tilde{\varepsilon}) \, G_{\text{MF}}(\varepsilon,\tilde{\varepsilon})\, , \label{omegaprod_eps}
\eeq
where $\tilde{\varepsilon} \to \varepsilon$ as $\varepsilon\to\varepsilon_c = 0$. One can thus write, as $\varepsilon\to\varepsilon_c$, 
\beq
\omega_{\text{MF}}(\varepsilon) \to \omega^{(1)}({\varepsilon}) \, g_{\text{MF}}(\varepsilon)\, , \label{omegaprod_epsc}
\eeq
where $g_{\text{MF}}(\varepsilon) = G_{\text{MF}}(\varepsilon,\varepsilon)$, for $\varepsilon \to \varepsilon_c$. Figure \ref{figure_eta} shows the function
\beq
\eta(\varepsilon) = h(\sqrt{-2\varepsilon}) = \varepsilon - \tilde{\varepsilon}\, ;
\label{eta} 
\eeq
as $\varepsilon \to \varepsilon_c = 0$, $\eta(\varepsilon) \propto  -\sqrt{-\varepsilon}$. Since $|\eta(\varepsilon)|$ is the difference between the energy $\tilde{\varepsilon}$ singled out by the saddle point and the energy $\varepsilon$ at which the density of states is calculated, it somehow measures also the ``distance'' between the function $G_{\text{MF}}(\varepsilon,\tilde{\varepsilon})$ and the function $g_{\text{MF}}(\varepsilon) = G_{\text{MF}}(\varepsilon,{\varepsilon})$. From Fig.\ \ref{figure_eta} we see that this difference reaches its maximum (roughly equal to $1.2\times 10^{-2}$) around the center of the energy density range. Comparing this value to the width of the energy range itself we see that this difference is at most of the order of 2\%.

\section{The one-dimensional $XY$ model}
\label{sec_1dXY}

Let us now consider the one-dimensional $XY$ model, which is a system of $N$ planar spins with nearest-neighbor coupling, described by the Hamiltonian
\beq
\mathcal{H}_{1d} = - \sum_{i=1}^{N-1} \cos \left(\vartheta_{i+1} - \vartheta_i \right)\, ,
\label{H1d}
\eeq
where, as in the mean-field $XY$ model, $\vartheta_i \in [0,2\pi)$, so that the configuration (or phase) space of the system is the torus $\mathbb{T}^N$. This model does not have a bulk broken symmetry phase; it is ordered only in its state of minimum energy. Hence, the phase transition from a ferromagnetic to a paramagnetic phase occurs at $\varepsilon_c = \varepsilon_{\text{min}} = -1$, and at zero temperature. It belongs to the class (\ref{H}) with $n=2$ and $J_{ij} = 1$ for $i$ and $j$ nearest-neighbors and zero otherwise.

As we shall see in the following, also for this model the density of states can be written as 
\beq
\omega_{1d}(\varepsilon) = \omega^{(1)}(\tilde{\varepsilon}) \, G_{1d}(\varepsilon,\tilde{\varepsilon})\, , \label{omegaprod_eps_1d}
\eeq
where, in this case, $\omega^{(1)}$ is the density of states of the one-dimensional Ising model
\beq
\mathcal{H}_{1d}^{(1)} = - \sum_{i=1}^{N-1} \sigma_{i}\sigma_{i+1} \, , 
\eeq
and  $\tilde{\varepsilon} \to \varepsilon$ as $\varepsilon\to\varepsilon_c = \varepsilon_{\text{min}}$. One can thus write, as $\varepsilon\to\varepsilon_c$, 
\beq
\omega_{1d}(\varepsilon) \to \omega^{(1)}({\varepsilon}) \, g_{\text{1d}}(\varepsilon)\, , \label{omegaprod_epsc_1d}
\eeq
where $g_{1d}(\varepsilon) = G_{1d}(\varepsilon,\varepsilon)$, for $\varepsilon \to \varepsilon_c$. The derivation follows very closely that of the mean-field model, with a few differences that will be underlined.

Let us fix $\vartheta_N = 0$, and leave open the boundary condition at the other side of the chain. As in the mean-field case, the Ising stationary configurations are those where the angles $\overline{\vartheta}$ are either $0$ or $\pi$. However, their energy is no longer parametrized by $N_\pi$. On an Ising stationary configuration, the energy can be written as 
\beq
\mathcal{H}_{1d}(\overline{\vartheta}_1,\ldots,\overline{\vartheta}_{N-1}) = \mathcal{H}_{1d}^{(1)} = 2N_d - N + 1~,
\eeq
where $N_d$ is the number of the domain walls in the configuration, i.e., the number of flips between $\overline{\vartheta} = 0$ and  $\overline{\vartheta} = \pi$ (and viceversa) along the chain. This implies that one can no longer use the definition (\ref{partition}) of the neighborhoods $U_p$ to build the partition of the configuration space, because this would imply that stationary points with the same energy would give different contributions.

Let us then change variables from $(\vartheta_1,\ldots,\vartheta_N)$ to $(x_1,\ldots,x_N)$ as follows:
\beq
\left\{ \begin{array}{llll}
x_k & = & \vartheta_{k+1} - \vartheta_k & \text{if} ~k=1,\ldots,N-1, \\
&& \\
x_N & = &\vartheta_N = 0\,.&
\end{array} \right.
\eeq
In the new variables the Ising stationary points are still such that $\overline{x}_k = 0$ or $\overline{x}_k = \pi$, but now the energy is given in terms of the number of $x$'s equal to $\pi$, because the number of domain walls $N_d$ is precisely that number. One can thus define the partition of the configuration space using the neighborhoods $U_p$ defined as 
\beq
U_p = \left\{
\begin{array}{ccc}
{\displaystyle x_i \in \left[\frac{\pi}{2},\frac{3\pi}{2} \right]}  & \text{if} &   \overline{x}_i = \pi  \\
& & \\
{\displaystyle x_i \in \left[\frac{3\pi}{2},\frac{\pi}{2} \right]}  &\text{if}&    \overline{x}_i = 0  
\end{array} \right .
\label{partition_1d}
\eeq
and write the density of states of the 1-$d$ $XY$ model as
\beq
\omega_{1d}(\varepsilon) = \sum_{N_d = 0}^{N-1} \nu(N_d)\, G_{1d}(\varepsilon,N_d)
\eeq
where 
\beq
\nu(N_d) = \frac{(N-1)!}{N_d! (N - N_d - 1)!} 
\eeq
is the number of Ising configurations with $N_d$ domain walls, i.e., the density of states $\omega^{(1)}(\varepsilon')$ of the one-dimensional Ising model with energy density
\beq 
\varepsilon' = \frac{2N_d - N + 1}{N}\, ,
\label{epsilon_nd}
\eeq
and 
\beq
G_{1d}(\varepsilon,N_d) = \int_{\pi/2}^{3\pi/2} dx_1\cdots dx_{N_d} \int_{3\pi/2}^{\pi/2} dx_{N_d + 1}\cdots dx_{N - 1}\, \delta \left(-\sum_{k=1}^{N-1} \cos x_k - N\varepsilon \right)\, . 
\label{G1d}
\eeq
The computation then proceeds following very closely what already done for the mean-field case. The 1-$d$ case is even simpler, because one can directly compute $G_{1d}$ as a function of the energy density, without the need to consider it as a function of the magnetization. Using the integral representation of the $\delta$ and integrating on the $x$ variables we can write in the large $N$ limit
\beq
G_{1d}(\varepsilon,n_d) = \frac{1}{2\pi} \int_{-\infty}^{\infty} dq \exp\left\{N \left[-iq\varepsilon + n_d \log b(q) + (1-n_d) \log a(q) \right]\right\} \, ,
\eeq
where $n_d = N_d/N$ and the functions $a$ and $b$ are given by 
\bea
a(q) & = & \int_{3\pi/2}^{\pi/2}dx \exp\left(-iq \cos x \right)\, , \\
b(q) & = & \int_{\pi/2}^{3\pi/2}dx \exp\left(-iq \cos x \right)\, . 
\eea
Performing again a saddle point with $q = -i\gamma$ we get, in the $N \to\infty$ limit,
\beq
G_{1d}(\varepsilon,n_d) = \frac{1}{\pi}  \exp\left\{N \left[-\gamma\varepsilon + n_d \log \tilde b(\gamma) + (1-n_d) \log \tilde a(\gamma) \right]\right\} \, ,
\label{G1saddle}
\eeq
where 
\bea
\tilde a(\gamma) & = & \tilde A(\gamma,0)\, , \\
\tilde b(\gamma) & = & \tilde B(\gamma,0)\, , 
\eea
with $\tilde A$ and $\tilde B$ given by Eqs.\ (\ref{Agamma}) and (\ref{Bgamma}), respectively, and where $\gamma$ satisfies the self-consistency equation
\beq
\varepsilon = (1 -n_d) \,\frac{I_1(\gamma)-L_{-1}(\gamma)}{I_0(\gamma)-L_0(\gamma)} + n_d\,\frac{I_1(\gamma)+L_{-1}(\gamma)}{I_0(\gamma)+L_0(\gamma)}\, .
\label{selfconsistency1d}
\eeq
We can thus realize that Eqs.\ (\ref{G1saddle}) and (\ref{selfconsistency1d}) coincide with the same equations derived for the mean-field case, i.e., Eqs.\ (\ref{GMFsaddle}) and (\ref{autoconsistenza_punto_sella}), provided 
\beq
\left\{
\begin{array}{ccl}
m & \to & \varepsilon  \\ 
n_\pi & \to & 1 - n_d
\end{array} \right.
\label{subs}
\eeq
The latter reflect the fact that in the 1-$d$ case the transition occurs at the minimum value of $\varepsilon$ instead of at the maximum. 

From now on, the calculation of $\omega_{1d}(\varepsilon)$ is exactly the same as that of $\omega_{\text{MF}}(m)$, with the substitutions (\ref{subs}). A given value of $\tilde n_d$ of $n_d$ will be singled out, which corresponds to an energy density $\tilde\varepsilon$ via Eq.\ (\ref{epsilon_nd}). We thus obtain
\beq
\omega_{1d}(\varepsilon) = \omega^{(1)}(\tilde{\varepsilon}) \, G_{1d}(\varepsilon,\tilde{\varepsilon})\, , 
\eeq
where $\tilde{\varepsilon} \to \varepsilon$ as $\varepsilon\to\varepsilon_c = \varepsilon_{\text{min}}$; more precisely, defining the function 
\beq
\zeta(\varepsilon) = \varepsilon - \tilde{\varepsilon} = h(m = \varepsilon + 1) \, ,
\label{zeta} 
\eeq
where $h(m)$ is the function (\ref{h}) defined for the mean-field $XY$ model, we have that $\zeta \to 0$ when $\varepsilon\to\varepsilon_c = \varepsilon_{\text{min}} = -1$, and in particular $\zeta\propto -(1 +\varepsilon)$ for $\varepsilon$ close to $\varepsilon_c = -1$. If one plots $\zeta$ as a function of $\varepsilon$ one thus obtains exactly the same curve reported in Fig.\ \ref{figure_h}, with the horizontal axis shifted so that $\varepsilon\in[-1,0]$. Since $|h(m)|$ is maximum for $m \simeq 0.75$, the function $|\zeta(\varepsilon)|$ reaches its maximum value (roughly equal to $0.15$) around $\varepsilon\simeq -0.25$; the maximum difference between $\varepsilon$ and $\tilde{\varepsilon}$ in this case is around 15\% of the full energy density range, larger than in the mean-field case. 

\section{Concluding remarks}
\label{sec_conclusions}
The present paper has been mainly devoted to discuss the validity of the relation (\ref{omega_appr}), put forward in \cite{prl2011}, in the special cases of the mean-field and $1$-$d$ $XY$ models. We have shown that a slightly more general formula, Eq.\ (\ref{omega_prod}), holds, which reduces to the previous one in the limit $\varepsilon \to \varepsilon_c$. 

The two models we have dealt with are very special and both of them are exactly solvable in the microcanonical ensemble. This feature is crucial for the derivation we have presented. As a consequence, a generalization of the computations to $O(n)$ models with short-ranged interactions on a $d$-dimensional lattice is not straightforward at all, the difficulties being similar to exactly solving their thermodynamics in the microcanonical ensemble. 

This notwithstanding, we can learn something from these results. 
The present work confirms that, as already noted in \cite{prl2011}, Eq.\ (\ref{omega_appr}) can not be exact for a generic value of the energy density; at most, it could be valid for $\varepsilon = \varepsilon_c$. Indeed, it was already pointed out in \cite{prl2011} that the relation (\ref{omega_appr}) could not be exact for a generic $O(n)$ model, since the specific heat critical exponent $\alpha$ of a $O(n)$ model would then have the correct sign, but the wrong absolute value. More precisely, Eq.\ (\ref{omega_appr}) implies that if $\alpha_I$ is the microcanonical specific heat exponent of the Ising model on a given lattice, then the microcanonical specific heat exponent of the $O(n)$ model on the same lattice and with the same interactions is  $\alpha=-\alpha_I$, regardless of $n$. In $d=3$, for instance, this yields the correct sign of the $O(n)$ exponents, because $\alpha_I > 0$ so that $\alpha < 0$; the $O(n)$ specific heat is not divergent, but cuspy at the transition. However, the absolute value of the exponent is wrong, because it should depend on $n$, as shown by well-established results for the $O(n)$ universality classes \cite{PelissettoVicari:physrep2002}. It is worth noting that, here and in the following, we are dealing with the specific heat critical exponents defined in the microcanonical ensemble: these are related to the usual critical exponents $\bar\alpha$ defined in the canonical ensemble by $\alpha = \bar{\alpha}/(1-\bar{\alpha})$ \cite{KastnerPrombergerHuller:jstatphys2000}, so that microcanonical results can be easily carried over to the canonical ensemble\footnote{In particular, if $\bar\alpha \in [-1,0]$ then  $\alpha \in \left[-\frac{1}{2},0\right]$; we note that the relation $\alpha = \bar{\alpha}/(1-\bar{\alpha})$ given in \cite{KastnerPrombergerHuller:jstatphys2000} holds for any $\bar\alpha < 1$.}. 

The result $\alpha = -\alpha_I$ follows from Eq.\ (\ref{omega_appr}) by assuming that the function $g^{(n)}(\varepsilon)$ is a generic function which does not contain any explicit information on  the phase transition, i.e., is analytic with a generic Taylor expansion. If we proceed in an analogous way assuming that Eq.\ (\ref{omega_prod}) holds for a generic $O(n)$ model, we still find the correct sign of the specific heat critical exponents as with Eq.\ \ref{omega_appr}, but we do no longer have any contradiction with the known results on the values of the exponents. Indeed, assuming that $G^{(n)}(x,y)$ is a generic (i.e., analytic) function because it should not contain any information about the phase transition, it can be shown that the critical exponent $\alpha$ of the continuous model can be any real number in $[-1,0)$. This range of values is in agreement with known results \cite{PelissettoVicari:physrep2002}; moreover, although it does not predict a precise value of $\alpha$, it still correctly implies that the specific heat of $O(n)$ lattice spin models does not diverge for $n >1$. The details about the predictions of Eqs. (\ref{omega_appr}) and (\ref{omega_prod}) as to the critical exponent $\alpha$ are reported in Appendix \ref{app}. The latter observations, together with the observed near-equality of critical energies  of $O(n)$ models defined on the same lattice at different $n$ \cite{prl2011}, suggest that the relation (\ref{omega_prod}) might have a more general validity than being restricted to the two special cases considered in this paper.

Although the above argument shows that a more general validity of Eq. (\ref{omega_prod}) as an exact result can no longer be excluded on the basis of the predictions for the exponent $\alpha$, it would imply, as in the case of  Eq.\ (\ref{omega_appr}) that was already discussed in \cite{prl2011}, a ``Patrascioiu-Seiler'' scenario in $d=2$, i.e.,  the presence of some kind of phase transition in two-dimensional $O(n)$ models also for $n >2$ \cite{PatrascioiuSeiler:prl1995,PatrascioiuSeiler:prb1996}. This scenario, although not ruled out by rigorous results, is believed to be unlikely on the basis of numerical simulations (see e.g.\ Ref.\ \cite{CaraccioloEtAl:prl1995,PatrascioiuSeiler:prl1996,CaraccioloEtAl:prl1996}). 

\acknowledgments

CN acknowledges support from the EGIDE scholarship funded by Minist\`ere des Affaires \'etrang\`eres (France).

\appendix
\section{Specific heat critical exponent}
\label{app}
In Sec.\ \ref{sec_conclusions}, we have discussed the implications of Eqs.\ (\ref{omega_appr}) and (\ref{omega_prod}) in case they would exactly hold. Here we give the details about the predictions on the specific heat critical exponent $\alpha$ obtained by assuming that the density of states has the form given by Eqs.\ (\ref{omega_appr}) or (\ref{omega_prod}), respectively. Let us recall that, in the microcanonical ensemble, the specific heat is defined as
\beq\label{cesare_2}
C(\varepsilon) = -\frac{\left[s'(\varepsilon)\right]^2}{s''(\varepsilon)}\,,
\eeq
where $s(\varepsilon)$ is the entropy density and the temperature is defined as $T(\varepsilon) = 1/s'(\varepsilon)$. With $s'(\varepsilon)$ and $s''(\varepsilon)$ we denote the first and second derivative of the function $s(\varepsilon)$.

Let us consider a short-range $O(n)$ model and assume the relation (\ref{omega_appr}) holds as an equality. We assume in the following that the phase transition occurs for a value of the energy density in the interior of the domain of the entropy density\footnote{As a consequence, what follows does not apply to the two models we considered in the bulk of the paper, the 1-$d$ and the mean-field $XY$ models.}.
Without loss of generality, let us shift the energy density $\varepsilon$ such that $\varepsilon_c=0$. The entropy density of the continuous model can then be written as:
\beq\label{cesare_1}
s(\varepsilon) = s_I(\varepsilon) + \log f(\varepsilon)\,,
\eeq
where here and in the following we use the notation $s_I(\varepsilon)$ instead of $s^{(1)}(\varepsilon)$ for the entropy density of the Ising model, to avoid possible misunderstanding with derivatives. We also omit the symbol ${(n)}$ indicating which $O(n)$ model we are considering because our arguments do not depend on it. Finally, we denoted $g^{(n)}(\varepsilon)^{1/N}$ by $f(\varepsilon)$.

Let us now consider, for the moment, only energy densities larger than the critical one, i.e., $\varepsilon > 0$. Three facts are relevant for the following:
\begin{enumerate}
\item we consider $0<\alpha_I<1$, i.e., the case $d>2$. Moreover, because the critical temperature of the Ising models is finite, $s''_I(\varepsilon)\propto \varepsilon^{\alpha_I}$ for $\varepsilon \to 0^+$.
\item $s'(\varepsilon)$ is finite around $\varepsilon = 0$ because the critical temperature of the continuous model does not vanish at the transition.
\item we assume $f(\varepsilon)$ is analytical, consistently with the discussion in Sec.\  \ref{sec_intro}. 
We can then expand $f(\varepsilon)$ in a Taylor series around $\varepsilon = 0$.
\end{enumerate}
Inserting Eq.\ (\ref{cesare_1}) into  Eq.\ (\ref{cesare_2}), we get
\beq\label{cesare_3}
C(\varepsilon) = -\frac{\left[s'_I(\varepsilon) + \frac{g'(\varepsilon)}{f(\varepsilon)}\right]^2}{s''_I(\varepsilon) + \frac{g''(\varepsilon)}{f(\varepsilon)} - \left[\frac{g'(\varepsilon)}{f(\varepsilon)}\right]^2}\,.
\eeq
Using the expansions described above around $\varepsilon = 0$, neglecting the higher order terms and expanding the fraction, we obtain
\beq\label{cesare_4}
C(\varepsilon) \simeq a_+ + b_+\, \varepsilon^{\alpha_I}\,\qquad (\varepsilon \to 0^+),
\eeq
where $a_+$ and $b_+$ are constants whose exact value is irrelevant to our purposes. We can repeat the same calculations for $\varepsilon<0$, obtaining the same result as in Eq.\ (\ref{cesare_4}) but for that $\varepsilon \to -\varepsilon$ and that the constants may be different. Hence the specific heat close to $\varepsilon = 0$ is 
\beq\label{cesare_4bis}
C(\varepsilon) \simeq a_\pm + b_\pm\, \left|\varepsilon \right|^{\alpha_I}\,.
\eeq
We then obtain the result stated in Sec.\ \ref{sec_conclusions}: the specific heat of the continuous model does not diverge at the transition and the critical exponent $\alpha$ of the continuous model is related to the one of the Ising model via $\alpha = -\alpha_I$.

With a similar reasoning we can also deal with the case in which we consider Eq.\ (\ref{omega_prod}) to be exact.
As before, we start by considering $\varepsilon > 0$. Assuming Eq.\ (\ref{omega_prod}) holds as an equality, the entropy density of the continuous model is
\beq\label{cesare_5}
s(\varepsilon) = s_I(\varepsilon) + f(\varepsilon,\tilde{\varepsilon}(\varepsilon))\,,
\eeq
where we denoted by $f(\varepsilon,\tilde{\varepsilon}(\varepsilon))$ the function $(1/N) \log g^{(n)}(\varepsilon,\tilde{\varepsilon}(\varepsilon))$. In this case, $f$ is a function of two variables: again, we assume it is analytic and expand it around $\varepsilon = 0$, such that
\beq\label{cesare_6}
f(x,y)\simeq f_0+f_1 x + f_2 y + f_3 xy + f_4 x^2 + f_5 y^2 + f_6 x^2y + f_7 x y^2 + f_8 x^3 + f_9 y^3\,,
\eeq
where $x$ and $y$ are shorthands for $\varepsilon$ and $\tilde{\varepsilon}$ and the $f_i$'s are constants whose exact value is irrelevant to our purposes. At variance with the previous case, $\tilde{\varepsilon}(\varepsilon)$ contains some information about the transition because it vanishes for $\varepsilon \to 0$; we should then admit the possibility of a singular dependence on $\varepsilon$, writing $\tilde{\varepsilon}(\varepsilon)\propto \varepsilon^{\theta}$ with $\theta > 0$ for $\varepsilon \to 0^+$. 

Using the information on the behavior of $s'' _I(\varepsilon)$ around $\varepsilon = 0$ and integrating two times, we get
\beq\label{cesare_7}
s_I(x)\simeq a_0+a_1 x +a_2 x^{\alpha_I+2}\,,
\eeq
where the $a_i$'s are suitable constants. Inserting Eqs.\ (\ref{cesare_6}) and (\ref{cesare_7}) into the equation for the entropy of the continuous model, Eq.\ (\ref{cesare_5}), we get:
\beq\label{cesare_8}
s(\varepsilon)\simeq a_0+a_1\varepsilon^{\theta} + a_2\varepsilon^{\theta(\alpha_I+2)} +f_0 + f_1 \varepsilon + f_2 \varepsilon^{\theta} + f_3 \varepsilon^{\theta+1} + f_4 \varepsilon^2 + f_5 \varepsilon^{2\theta} + f_6 \varepsilon^{2+\theta} + f_7 \varepsilon^{1+2\theta} +f_8 \varepsilon^3 +f_9 \varepsilon^{3\theta}\,.
\eeq
Taking the first and the second derivative of the previous expression and renaming the constants, we obtain
\beq\label{cesare_9}
s'(\varepsilon)\simeq b_1 \varepsilon^{\theta-1} +b_2 \varepsilon^{\theta(\alpha_I+2)-1} +h_1+h_2\varepsilon^{\theta-1} + h_3\varepsilon^{\theta} + h_4 \varepsilon + h_5\varepsilon^{2\theta-1}  + h_6 \varepsilon^{\theta+1} + h_7 \varepsilon^{2\theta} + h_8 \varepsilon^2 + h_9 \varepsilon^{3\theta -1}\,,
\eeq
and
\beq\label{cesare_10}
s''(\varepsilon)\simeq c_1 \varepsilon^{\theta-2} + m_3 \varepsilon^{\theta-1} + m_4 + m_5 \varepsilon^{2\theta -2} + m_6 \varepsilon^{\theta} + m_7 \varepsilon^{2\theta -1} + m_8 \varepsilon + m_9 \varepsilon^{3\theta-2}\,.
\eeq
The quantity $\theta$ is unknown. However, since the specific heat of the continuous model does not vanish at the transition, the above expressions imply the constraint $\theta \geq 2$. Moreover, if $\theta >3$, the linear term in Eq.\ (\ref{cesare_10}) would dominate. Hence the range of values for $\theta$ to be considered is $\theta \in (2,3]$; if $\theta >3$ or $\theta=2$, the leading behavior of $s''(\varepsilon)$ would be the same as that given by Eq.\ (\ref{cesare_10}) with $\theta=3$. 

The leading behavior of Eqs.\ (\ref{cesare_9}) and (\ref{cesare_10}) is then $s'(\varepsilon)\simeq h_1+h_4 \varepsilon$ and $s''(\varepsilon)\simeq m_4+c_1\varepsilon ^{\theta -2}$. Inserting these results into the expression (\ref{cesare_2}) for the specific heat, we obtain
\beq\label{cesare_11}
C(\varepsilon)\simeq -\frac{\left(h_1+h_4 \varepsilon \right)^2}{m_4+c_1\varepsilon^{\theta-2}}\simeq c_+ + d_+\,\varepsilon^{\theta-2}\, \qquad (\varepsilon \to 0^+)\,.
\eeq
Repeating the same calculations for $\varepsilon < 0$ and combining the result with Eq.\ (\ref{cesare_11}) we obtain the behavior of the specific heat close to the transition,
\beq\label{cesare_11_bis}
C(\varepsilon)\simeq c_\pm + d_\pm\,\left| \varepsilon \right|^{\theta-2}\,.
\eeq
The above expression, together with the above bounds on $\theta$, shows that the specific heat of the continuous model does not diverge and its critical exponent $\alpha$ is determined by $\theta$, which is model dependent. Varying $\theta$ in its allowed range we obtain $\alpha \in [-1,0)$.

\bibliography{/Users/casetti/Work/Scripta/papers/bib/mybiblio,/Users/casetti/Work/Scripta/papers/bib/statmech}

\end{document}